\documentclass[preprint,aps,amsmath,byrevtex,prd,titlepage,
nofootinbib]{revtex4-1}

\usepackage{dcolumn}
\usepackage{amsmath}
\usepackage{amssymb}
\usepackage{bm}
\usepackage{hyperref}
\usepackage{graphicx}
\usepackage{slashed}

\newcommand{\beq}{\begin{equation}}
\newcommand{\eeq}{\end{equation}}
\newcommand{\eq}[1]{Eq.~(\ref{#1})}

\begin{document}

\title {On Some Recent Ideas on the Proton Radius Puzzle and Lepton Anomalous Magnetic Moments}
\author {Michael I. Eides}
\email[Email address: ]{eides@pa.uky.edu, eides@thd.pnpi.spb.ru}
\affiliation{Department of Physics and Astronomy,
University of Kentucky, Lexington, KY 40506, USA\\
and Petersburg Nuclear Physics Institute,
Gatchina, St.Petersburg 188300, Russia}


\begin{abstract}
We discuss recent suggestions on the resolution of the proton radius puzzle put forward in \cite{gs2013,rob2013} and discovery of a nonperturbative quantum-electrodynamic contribution of order $(\alpha/\pi)^5$ to lepton anomalous magnetic moments  announced in \cite{mish2013,fapas2014}. We demonstrate that the ideas of \cite{gs2013,rob2013} do not resolve the proton radius puzzle. We explain why the nonperturbative correction calculated in \cite{mish2013,fapas2014} does not exist.

\end{abstract}


\preprint{UK/13-05}

\maketitle

\section{Introduction}

Search for new physics beyond the Standard Model in laboratory experiments is perhaps one of the most interesting avenues of research in modern fundamental physics. Unfortunately despite the spectacular discovery of the long awaited Higgs boson, thus far high energy experiments failed to produce convincing evidence of new physics. In these circumstances a lot of attention attracted recently low energy experiments and high accuracy theoretical calculations that also can provide a window on new physics. Below we discuss some recent ideas on resolution of the proton radius puzzle
and the electron anomalous moment calculations.

\section{Proton Radius}

It is now more than three years since the measurement of the Lamb shift in muonic hydrogen at the PSI \cite{pohletal} opened the Pandora box of the proton radius puzzle. The essence of the problem is a discrepancy between the values of the proton radius extracted from the electron-proton scattering and measurements of the Lamb shift in ordinary hydrogen on the one hand, and the value of the proton radius extracted from the measurement of the Lamb shift in muonic hydrogen on the other hand. The discrepancy is as large as seven sigma, see discussion in \cite{mtn2012}.

There were numerous attempts to find a flaw either in the electron-proton scattering theory or in the Lamb shift theory, see review in \cite{rgmp2013}. It is fair to say  that the proton radius puzzle successfully withstood all theoretical attacks and due to the latest experimental results \cite{antogn2013} is now even more acute than ever.

A new attack on the proton radius puzzle is presented in \cite{gs2013}. The idea of the authors is that different definitions of the proton radius are used in the discussion of the electron-proton scattering experiments and in the Lamb shift theory. Without any justification it is assumed in \cite{gs2013} that the proton radius used in calculations with the muonic hydrogen does not coincide with the proton radius defined in a relativistic invariant way via the slope of the electric Sachs form factor. Use of different definitions surely could explain the discrepancy between the values of the proton radius extracted from the scattering experiments on the one hand and from the muonic hydrogen measurements on the other hand. The only problem with this explanation is that it is based on misunderstanding. The proton radius used in the Lamb shift theory is defined in a relativistic invariant way as the slope of the electric Sachs form factor in the expansion over the four-momentum squared, see, e.g., discussion in \cite{egsrev,egsmon2006}. Exactly this proton radius is extracted from the experimental results on the electron-proton scattering. Therefore, the ideas put forward in \cite{gs2013}, as well as in an earlier \cite{rob2013} do not resolve the proton radius puzzle.

\section{Positronium Singularities of the Polarization Operator and Lepton Anomalous Magnetic Moments}

Calculation of the QED corrections to lepton anomalous magnetic moments is the classical playground of perturbation theory. In fact, the very first prediction of the newly minted QED was the Schwinger's calculation of the one-loop correction to the electron anomalous magnetic moment (AMM) \cite{schw1948}. Nowadays all QED corrections to the electron AMM up to and including the five-loop ($(\alpha/\pi)^5$) contributions are calculated, see the review \cite{ahkn2012}.

It was claimed recently \cite{mish2013,fapas2014} that the perturbative calculation of the QED contributions to lepton AMMs is incomplete, and should be amended by the account for a "nonperturbative" effect connected with the positronium poles in the photon polarization operator. According to \cite{mish2013,fapas2014} these positronium poles in the exact polarization operator generate a new contribution of order $(\alpha/\pi)^5$ that is missed in the standard perturbative treatment and that should be added to the perturbative one. If correct, the conclusion about the missing positronium poles would affect many other high-order QED calculations.

The problem of account for the positronium poles in the polarization operator in low energy QED calculations was addressed, exhaustively explored and solved long time ago in \cite{braun1967} (see also \cite{bcr1973}). It was proved that  almost in all low energy perturbative calculations, when the momentum $k$ flowing through the polarization operator is far from the positronium singularities $|k^2-4m^2|>\alpha^2m^2$, perturbative contributions provide complete QED result. Inclusion of positronium poles in this situation means an ill defined double counting. The only exception to this rule is the case when the polarization operator enters the diagram in the vicinity of the positronium poles $|k^2-4m^2|\leq\alpha^2m^2$. Then account for the positronium singularities becomes mandatory. An example of such situation is provided by the vacuum polarization insertion in the diagram with the virtual one-photon annihilation of positronium. The contribution of the positronium singularities to the positronium hyperfine splitting in this case was calculated in \cite{bcr1973}.

\begin{figure}[htb]
\includegraphics
[height=3cm]
{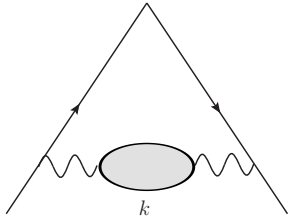}
\caption{\label{polamm}
Polarization Operator Contribution to AMM}
\end{figure}

Below we will consider in more detail calculation of the polarization operator contribution in Fig.~\ref{polamm} and demonstrate that the standard perturbative contributions give the complete QED contribution to AMM. Nothing we will say below is new and not contained in \cite{braun1967,bcr1973}, but the claims made in \cite{mish2013,fapas2014} show that the old arguments deserve to be repeated.

We start with a qualitative discussion. Recall how the formation of a QED bound state is described in the diagrammatic language. Consider an electron-positron four-point Green function (the polarization operator is just a special case of this four-point function). This Green function has a series of poles in the $s$-channel ($s=k^2$) corresponding to the energy levels of the electron-positron bound states -- levels of positronium of the form $Z_n(\alpha)/(s-E_n^2)$, $E_n=2m-\alpha^2m/(4n^2)$. These poles can be considered as singularities (poles) in $\alpha$, and an expansion in powers of $\alpha$ is clearly illegitimate in the vicinity of the poles, when $|s-4m^2|\leq m^2\alpha^2$. Besides bound states the singularities in $\alpha$ are due also to the continuum scattering states near the threshold,   $s- 4m^2\leq m^2\alpha^2$, see \cite{braun1967,bcr1973}. It is well known that these singularities arise as a result of summation of an infinite series of the perturbation theory diagrams, see, e.g., \cite{blp}.  Consider the diagrams with the (Coulomb) photon exchanges between the electron and positron lines that are responsible for the formation of the bound state.  For the generic external momenta each extra exchanged photon carries an extra suppression factor $\alpha\sim1/137$, and the diagrams with a large number of photon exchanges are strongly suppressed.  How is it possible that the sum of these diagrams develops poles at $s\sim 4m^2$ corresponding to the positronium bound states? The only way how it happens is that the series diverges  at $s\sim 4m^2$, and all terms in the series are of the same order in the vicinity of $s\sim 4m^2$. At first glance this cannot happen since, as we just explained, addition of an extra rung to the ladder of exchanged photons results in an extra small factor $\alpha$. But not so fast, this simple minded consideration completely ignores kinematic dependence of the graphs. A simple calculation shows that at $s\sim 4m^2$ each extra photon is accompanied by an extra factor $1/v\sim 1/\alpha$ ($v=\sqrt{1-4m^2/s}$). As a result all diagrams are of the same order and the series diverges at $s\sim 4m^2$. We can easily sum the infinite series of diagrams with the Coulomb exchanges far from the threshold, and check that the sum really develops poles at the positions of the positronium energy levels. For our goals it is important to emphasize that the positronium poles in the full polarization operator arise as a result of summation of an infinite series of one-particle irreducible perturbation theory diagrams, and these poles arise at the specific values of the four-momentum flowing through these perturbation theory diagrams.

Now we are ready to return to the problem of accounting for many Coulomb exchanges inside the polarization operator in the diagram in Fig.~\ref{polamm}. Let us consider contribution to AMM provided by the diagram in Fig. \ref{polamm} where the blob is substituted by any finite order perturbation theory diagram for the polarization operator. It is well known that the structure of singularities in the complex $k_0$ plane is such that it allows the Wick rotation, see, e.g., \cite{blp,pesschr}. After the Wick rotation the loop (and polarization operator) momentum is spacelike, $k^2=-k_E^2<0$. Simply from dimensional considerations we know that the characteristic integration loop momentum in this diagram is determined by the electron mass, and the dominant contribution to the diagram is produced by the region of integration momenta where Euclidean momenta $k_E^2\leq m^2$. This is true for calculation of the leading Schwinger contribution \cite{schw1948} that is reproduced in any decent text on quantum field theory, see, e.g, \cite{blp,pesschr}. Insertion of an one-particle irreducible polarization operator diagram with $n$ photon exchanges in the photon line in Fig. \ref{polamm} changes large momentum behavior of the photon propagator by the factor $\sim\ln(k_E^2/m^2)$ that does not change dominant integration momentum region. We see that the characteristic spacelike integration momenta $k^2<0$ are separated by a large gap $\sim 4m^2$ from the position of the positronium singularities at timelike momenta $k^2\sim 4m^2$. In this kinematics each polarization operator diagram with an extra Coulomb exchange introduces an extra power of $\alpha$ and to achieve the desired accuracy in calculation of AMM it is sufficient to consider only a finite number of lower order perturbation theory diagrams. All higher order diagrams with a larger number of exchanged photons are suppressed by powers of $\alpha$ and can be safely neglected\footnote{We do not discuss insertions of one particles reducible polarization operators that lead to the well known renormalon singularity \cite{lautr1977}.}. The positronium poles are nothing but the sum of the perturbation theory diagrams at $k^2\sim 4m^2$, and we see that for kinematical reasons they do not arise in calculation of the corrections to AMM.

The discussion above can be illustrated with the help of a toy model suggested in \cite{mish2013}. Consider the function $f(x)={a}/({x-a})$ (in the context of the present discussion one can think that $x$ is something like $k^2-4m^2$, and  $a$ is something like $-\alpha^2m^2/n^2$). Expanding this function in a power series and removing from this expansion, say, first ten terms

\beq
f(x)=
\frac{a}{x} +\frac{a^2}{x^2}+\dots+\frac{a^{10}}{x^{10}}
+\frac{a^{11}}{x^{11}}\frac{a}{x-a},
\eeq

\noindent
we observe that the remainder function $f_1(x)=(a^{11}/x^{11}){a}/({x-a})$ still has the pole with the same residue as the original function $f(x)$. This observation was used in \cite{mish2013} as an argument demonstrating that the pole contribution of the function $f_1(x)$ has to be accounted for together with the first few terms in the power series expansion. It was  claimed that since the residue of the function $f_1(x)$ at $x=a$ coincides with the residue of the function $f(x)$ the pole contribution of $f_1(x)$ is not suppressed in comparison with the perturbation theory contributions. However, as explained above, we are interested in the value of the integral of this series expansion with a certain weight, and the weight function chooses the values of the argument that are far from the pole. For illustrative purposes let us choose the weight in the form $\delta(x-b)$, where $|b-a|\gg a$. Then

\beq
f(b)=\int dx f(x)\delta(x-b)
=\frac{a}{b} +\frac{a^2}{b^2}+\dots+\frac{a^{10}}{b^{10}}+f_1(b),
\eeq

\noindent
where $f_1(b)=(a^{11}/b^{11})[a/(b-a)]$. We see that the contribution of the "pole" term in $f_1(b)$ is purely perturbative in the small parameter $a/b$, and addition to the sum of a few first terms of something like $f(b)=a/(b-a)$ as is effectively done in \cite{mish2013,fapas2014} is unjustified. Of course, this toy model is just an illustration of the considerations above.

To summarize, inclusion of the positronium poles (one pole in \cite{mish2013}, a series of poles in \cite{fapas2014}) on par with a few perturbation theory diagrams is an uncontrollable approximation that leads to wrong results. The positronium poles in the polarization operator arise from summation of the diagrams with any number of the Coulomb exchanges. The poles show up as divergences of this series at specific values of momentum. Necessity for account for the positronium poles arises when all terms in the series are of the same order and the series diverges. In other words the perturbation theory itself prompts when it becomes insufficient. This is not the case for the corrections to AMM, and it makes no sense to add positronium  pole terms to the perturbation theory contributions as is done in \cite{mish2013,fapas2014}.

\begin{figure}[htb]
\includegraphics
[height=6cm]
{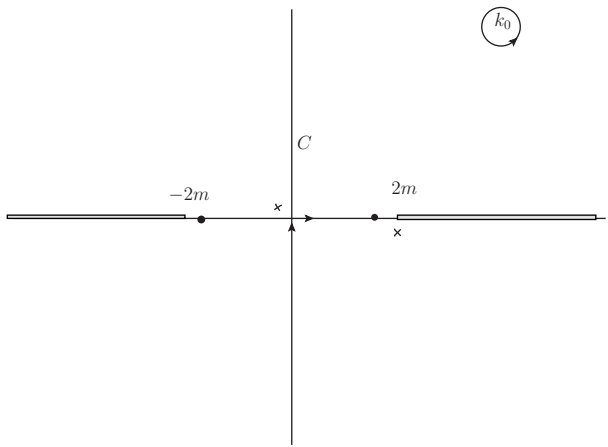}
\caption{\label{sing}
Singularities in the complex $k_0$ plane}
\end{figure}

One can prove purely perturbative nature of the vacuum polarization corrections to AMM from a slightly different perspective \cite{braun1967}. The diagram in Fig. \ref{polamm} with an exact one-particle irreducible polarization operator contains all polarization operator contributions to AMM. The exact polarization operator has, of course, positronium poles. Let us show that in the diagram in Fig. \ref{polamm} the contribution of the exact polarization operator reduces to a sum of contributions of a few lower order perturbation theory diagrams. Consider the loop integration over the loop momentum $k$ that coincides with the photon momentum. The singularities of the integrand in the complex plane $k_0$ in Fig. \ref{sing} are due to the polarization operator singularities (poles (bold dots) at $k^2=E_n^2=4m^2-\alpha^2m^2/n^2+\alpha^4m^2/(16n^2)$ and cuts at $k^2\geq 4m^2$). There are also singularities that arise  due to other propagators in the triangle diagram (crosses in Fig. \ref{sing}). We see that, as usual in the diagrammatic calculations, the structure of singularities allows rotation of the integration contour to the imaginary axis. After this rotation the contour is at a distance about $2m$ from the positronium singularities. As we discussed above for such momenta the exact polarization operator is a sum of a convergent perturbation theory expansion in $\alpha$, what proves validity of perturbation theory for calculation of radiative corrections to AMM.

Notice that in some other cases account for the positronium bound states is necessary in QED calculations. This happens when the singularities that depend on external momentum pin the integration contour to the positronium singularity, or when the polarization operator for kinematical reasons should be calculated near the positronium singularities. An example of such situation is provided by the contribution of the one-photon annihilation diagram to the positronium hyperfine splitting. The proper way to treat this problem was discovered and exhaustively discussed in \cite{bcr1973}.

For completeness let us consider also another argument in favor of account for the positronium poles in AMM calculations put forward in \cite{mish2013,fapas2014}. It is based on the well known dispersion relation representation of the polarization operator  contribution to AMM

\beq \label{disprel}
\Delta a=\frac{\alpha}{\pi^2}\int \frac{ds}{s} \mbox{Im} \Pi(s+i\epsilon)K(s),
\eeq

\noindent
where

\beq
K(s)=\int_0^1 dx\frac{x^2(1-x)}{x^2+(1-x)\frac{s}{m^2}}.
\eeq

It is claimed in \cite{mish2013,fapas2014} that considering only the perturbative contributions to the imaginary part in \eq{disprel} one misses the positronium pole contributions. Let us consider this argument in more detail, and apply it to the dispersion relation for the polarization operator itself. It is clear that each perturbation theory contribution to the polarization operator can be restored from its imaginary part with the help of the dispersion relation. Summing these perturbation theory contributions to the polarization operator we restore the total polarization operator that contains positronium poles as we explained above. But notice that restoring this polarization operator via dispersion relations for the perturbation theory diagrams we never encountered the positronium poles. Is there a contradiction? No, we just  observed that the sum of dispersion integrals of perturbative imaginary parts does not coincide with dispersion integral of the sum of imaginary parts. The summation and integration are in this case noncommutative as was first discovered in \cite{braun1967} (see also \cite{bcr1973}). The case of the dispersion integral in \eq{disprel} is similar to the case of the dispersion integral for the polarization operator itself, and we should not include positronium singularities in this dispersion relation on par with the perturbation theory contributions for the same reasons. Of course, if we knew the exact expression for the imaginary part of the full polarization operator we would be able to calculate the total polarization operator contribution to AMM with the help of \eq{disprel}. Since we do not know the exact polarization operator or its imaginary part,  perturbation theory remains the only practical way of calculating the QED corrections to AMM.

\section{Conclusions}

We have demonstrated that the ideas put forward in \cite{gs2013,rob2013} do not lead to resolution of the proton radius puzzle.  We have proved that the new nonperturbative correction \cite{mish2013,fapas2014} of order $(\alpha/\pi)^5$ to the electron and muon anomalous magnetic moments does not exist.

\acknowledgments

I am grateful to Dr. Keh-Fei Liu who attracted my attention to the discussion of the proton radius in \cite{gs2013}. This work was supported by the NSF grant PHY-1066054.

\end{document}